\documentclass[final,3p,times,twocolumn,a4]{elsarticle}

\usepackage{hyperref}

\def\bea{\begin{eqnarray}}
\def\eea{\end{eqnarray}}

\def\sp{\kern +3pt}
\def\ba{\begin{eqnarray}}
\def\ea{\end{eqnarray}}

\usepackage{graphics}
\usepackage{graphicx}
\usepackage{epsf} 
\usepackage{amsmath}
\usepackage{amssymb}
\usepackage{slashed}
\usepackage{bm}

\def\sfrac#1#2{{\textstyle \frac{#1}{#2}}}

\journal{Physics Letters B}

\begin{document} 

\begin{frontmatter}




\title{Improved empirical parametrizations of 
the $\gamma^\ast N \to N(1535)$ transition amplitudes 
and the  Siegert's theorem}

\author{G.~Ramalho  \vspace{-0.05in}}

\address{International Institute of Physics, Federal 
University of Rio Grande do Norte, 
Campus Lagoa Nova - Anel Vi\'ario da UFRN, 
Lagoa Nova, Natal-RN, 59070-405, Brazil \vspace{-0.35in}}



\begin{abstract}
Some empirical parametrizations of the 
$\gamma^\ast N \to N(1535)$ transition amplitudes 
violate the Siegert's theorem, that relates 
the longitudinal and the transverse amplitudes
in the pseudo-threshold limit 
(nucleon and resonance at rest).
In the case of the electromagnetic transition from   
the nucleon (mass $M$) to the resonance $N(1525)$ (mass $M_R$), 
the Siegert's theorem is sometimes expressed by the relation 
$|{\bf q}|   A_{1/2}= \lambda S_{1/2}$ 
in the pseudo-threshold limit, 
when the photon momentum $|{\bf q}|$ vanishes, 
and  $\lambda = \sqrt{2} (M_R -M)$.
In this article, we argue that the Siegert's theorem 
should be expressed by the relation 
$A_{1/2}  = \lambda S_{1/2}/|{\bf q}|$, 
in the limit $|{\bf q}| \to 0$. 
This result is a consequence 
of the relation $S_{1/2} \propto |{\bf q}|$,
when $|{\bf q}| \to 0$, as suggested 
by the analysis of the transition form factors 
and by the orthogonality between the nucleon 
and $N(1535)$ states.
We propose then new empirical parametrizations for the 
$\gamma^\ast N \to N(1535)$ helicity amplitudes, 
that are consistent with the data and the Siegert's theorem.
The proposed parametrizations follow 
closely the  MAID2007 parametrization, 
except for a small deviation in the amplitudes 
$A_{1/2}$ and $S_{1/2}$  when $Q^2 < 1.5$ GeV$^2$.
\end{abstract}


\end{frontmatter}

\section{Introduction}

The information relative to the structure 
of the electromagnetic transitions 
between the nucleon and the nucleon excitations 
($\gamma^\ast N \to N^\ast$) has been 
parametrized using different forms~\cite{NSTAR,Aznauryan12}.
The representations in terms of helicity amplitudes,
longitudinal and transverse,  
can be defined independently of the 
proprieties of the resonances.
Alternatively, one can use a representation in terms 
of structure form factors, that emphasize precisely the symmetries 
associated with the nucleon resonances.
The helicity amplitudes and the structure form factors 
are functions of the transition four-momentum 
transfer ($q$) squared, $q^2$, but are often represented 
in terms of $Q^2=- q^2$,
particularly in nucleon electroexcitation reactions ($Q^2 > 0$).
In general the different helicity amplitudes
are independent functions, 
except in some specific limits.
The same holds for the form factors.

Taking the case of the nucleon as example:
the electric and the magnetic form factors, 
$G_E$ and $G_M$, are independent functions,
except in the threshold limit, $Q^2= -4 M^2$,
where $G_E=G_M$ 
(threshold of the $\gamma^\ast \to N \bar N$ reaction).   
In the case of the $\gamma^\ast N \to N^\ast$ transitions, 
there are constraints between helicity amplitudes, 
or between form factors, at the pseudo-threshold limit.
The  pseudo-threshold limit is the 
limit where the photon momentum 
$|{\bf q}|$ vanishes, and both particles,
the nucleon ($N$) and the resonance, 
labeled here in general as $R$, are at rest.
In the  pseudo-threshold 
$Q^2= Q_{PS}^2 = -(M_R-M)^2$~\cite{Bjorken66,Devenish76}.

The condition that expresses the relation between 
different amplitudes (or form factors) at 
the pseudo-threshold is usually referred 
as the Siegert's theorem.
The Siegert's theorem was introduced first 
in studies related with nuclear physics~\cite{Bjorken66,Drechsel92}
and was later used in pion electroproduction 
reactions~\cite{AmaldiBook,Tiator2006,Drechsel2007,Tiator16}.

In this work, we study in particular 
the constraints of the  Siegert's theorem
in the $\gamma^\ast N \to N(1535)$ transition,
where $N(1535)$ is a spin $\frac{1}{2}$ state with negative parity 
($J^P=\frac{1}{2}^-$).
We will show in particular that some parametrizations 
of the $\gamma^\ast N \to N(1535)$ transition amplitudes, 
like the MAID2007 parametrization~\cite{Drechsel2007,Tiator16,MAID2009},  
are not consistent with the Siegert's theorem.
In order to grant that the Siegert's theorem 
is valid, one needs to ensure that $S_{1/2} \propto |{\bf q}|$,
near $|{\bf q}|=0$.
In the present article, we propose then new parametrizations 
for the amplitudes $A_{1/2}$ and $S_{1/2}$,
that are  consistent with both, the empirical data 
and the Siegert's theorem. 

The consequences of the Siegert's theorem for the 
$\gamma^\ast N \to \Delta(1232)$ 
and $\gamma^\ast N \to N(1520)$ helicity amplitudes
are discussed in a separate article~\cite{newPaper}.

\section{Siegert's theorem}

The parametrization of the current associated 
with a transition between the nucleon (state $J^P= \sfrac{1}{2}^+$)
and a $J^P= \sfrac{1}{2}^-$ resonance can be represented 
in terms of two form factors, 
$h_1$ and $h_3$ according with Ref.~\cite{Devenish76}.
At the  pseudo-threshold those form factors are related 
by the condition~\cite{Devenish76}
\ba
h_3 (Q_{PS}^2)=
\frac{M_R-M}{2M_R} h_1 (Q_{PS}^2).
\label{eqH1H3}
\ea
The functions $h_1, h_3$ can be related 
with the helicity amplitudes by 
$h_1 = - \sqrt{2} S_{1/2} /(|{\bf q}| b)$
and $h_3 = -  A_{1/2} /(M_R b)$,
where 
$b= e \sqrt{\frac{(M_R+ M)^2 + Q^2}{8 M (M_R^2 - M^2)}}$
and $e$ is the elementary electric charge.
The helicity amplitudes $A_{1/2}$ (transverse) 
and $S_{1/2}$ (longitudinal) 
will be defined precisely later [see Eqs.~(\ref{eqA12})-(\ref{eqS12})].

A direct consequence of the Eq.~(\ref{eqH1H3}) 
is 
\ba
A_{1/2} 
= \lambda \frac{S_{1/2}}{|{\bf q}|} 
\hspace{.6cm} 
(|{\bf q}| \to 0),
\label{eqSiegert}
\ea  
where we define 
\ba
\lambda = \sqrt{2}(M_R-M).
\ea
Note, that, we chose to include the ratio $S_{1/2}/|{\bf q}|$
in the previous relation.
In the case $|{\bf q}|=0$, the factor
$S_{1/2}/|{\bf q}|$ is interpreted as the limit $|{\bf q}| \to 0$.
This point is important, since it is assumed 
that $A_{1/2}$ and  $S_{1/2}/|{\bf q}|$ have the same order 
in $|{\bf q}|$, for small values of  $|{\bf q}|$. 
The consequence of this observation 
is that if $A_{1/2} = {\cal O}(1)$, 
meaning that $A_{1/2}$ converges to a constant in 
the pseudo-threshold limit, one can write 
also  $S_{1/2} = {\cal O}(|{\bf q}|)$, near $|{\bf q}| = 0$.

In this article, we will assume then, that, the amplitudes 
$A_{1/2}$ and $S_{1/2}$ behave, near the pseudo-threshold, as 
\ba
A_{1/2} = {\cal O}(1), \hspace{1cm} 
S_{1/2} = {\cal O}(|{\bf q}|).
\label{eqUseful}
\ea
The structure given by  Eqs.~(\ref{eqUseful}),
near the pseudo-threshold can be derived from 
the analysis of the multipole transition 
amplitudes~\cite{Bjorken66,AmaldiBook,Drechsel92,Tiator2006,Drechsel2007}.

In order to understand the meaning of the second relation
in (\ref{eqUseful}), we look for 
the charge density operator, $J^0$ 
(zero component of the transition current), 
in the pseudo-threshold limit.
The charge operator can be defined 
in terms of the Dirac ($F_1$) and Pauli ($F_2$) form factors 
[see Eq.~(\ref{eqJmu})].
When $J^0$ is projected into the spin states, 
which we represent by $\left< J^0 \right>$,
at the resonance rest frame,  one obtains 
\ba
\left< J^0 \right> = 
\tilde F_1 \left(\bar u_R \gamma_5 u \right), 
\ea 
where $u_R$ ($u$) is the Dirac spinor
of the resonance (nucleon) and 
\ba
\tilde F_1= F_1 + \eta F_2, 
\label{eqF1til}
\ea
with 
\ba
\eta= \frac{M_R-M}{M_R + M}.
\label{eqETA}
\ea
%
%
In the case where the initial and final 
state have the same spin projection, 
we can conclude, that,
in the pseudo-threshold limit at the $R$ rest frame: 
$\left(\bar u_R \gamma_5 u \right) \propto |{\bf q}|$.
Thus
\ba
\left< J^0 \right>  \propto \tilde F_1  |{\bf q}|.
\label{eqJ0a}
\ea
The previous condition defines the orthogonality 
between the nucleon and the resonance states
when $\left< J^0 \right> \to 0$,
which implies that $\tilde F_1 = {\cal O} (1)$,
($\tilde F_1 \to$ constant) or that 
$\tilde F_1$ scales with some power of $|{\bf q}|$,
in the pseudo-threshold limit.
The orthogonality between states at the pseudo-threshold
generalizes the nonrelativistic definition 
of orthogonality between states with different 
masses when the recoil (and the mass difference)
is neglected ($Q^2= -{\bf q}^2=0$).

Since the amplitude $S_{1/2}$ can also be defined by $J^0$, 
assuming current conservation\footnote{In 
the case of current conservation, 
the amplitude $S_{1/2}$ 
can be calculated using the operator 
$(\epsilon_0 \cdot J)|{\bf q}|/Q$, 
as in Eq.~(\ref{eqS12}), or using the operator $J^0$.}, 
in the cases where the spin 
projections are conserved 
(photon with zero spin projection),
we can also write $\left< J^0 \right> \propto S_{1/2}$.
Combined this result with the result (\ref{eqJ0a}), 
we conclude, that the orthogonality between the states, 
defined at the pseudo-threshold, implies 
\ba
S_{1/2} \propto \tilde F_1 |{\bf q}|.
\label{eqS12b}
\ea

In the following, we will also show that 
the first condition in (\ref{eqUseful}), 
$A_{1/2} = {\cal O}(1)$, implies that 
$\tilde F_1 = {\cal O}(1)$. 
Therefore, the combination of the result 
(\ref{eqS12b}) and $A_{1/2} \propto \tilde F_1$,
is compatible with the Siegert's theorem 
(\ref{eqSiegert}), apart from normalization factors.
To prove the relation (\ref{eqSiegert}), 
we need to look for the explicit 
parametrization of the 
amplitudes $A_{1/2}$ and $S_{1/2}$.

We introduce next the formalism associated 
with the electromagnetic transition current,
the electromagnetic form factors 
and the helicity amplitudes in 
the $\gamma^\ast N \to N(1535)$ transition.
Later, we discuss the implications 
of the Siegert's theorem in the structure 
of the transition form factors.

\section{$\gamma^\ast N \to N(1535)$ transition}

The $\gamma^\ast N \to N(1535)$ transition
can be represented, 
omitting the asymptotic states, 
in the units of 
the elementary electric charge $e$,   
as~\cite{Aznauryan12,N1535,S11scaling} 
\ba
J^\mu =
F_1 (Q^2) \left(\gamma^\mu - \frac{{\not \! q q^\mu}}{q^2} 
\right)\gamma_5
 + F_2(Q^2) \frac{i \sigma^{\mu \nu} q_\nu}{M_R + M} \gamma_5,
\nonumber \\
\label{eqJmu}
\ea 
where $F_1$ and $F_2$ are respectively 
the Dirac and Pauli form factors, 
as mentioned before.
Given the structure of Eq.~(\ref{eqJmu}), 
we can ensure, that, both components of 
the current, the Dirac and the Pauli terms,
are conserved separately.

\subsection{Helicity amplitudes (at the $R$ rest frame)}

Since the transition  $\gamma^\ast N \to N(1535)$ 
correspond to a transition between two states 
with spin $\frac{1}{2}$ (transition $\frac{1}{2}^+ \to \frac{1}{2}^-$), 
there are only two helicity amplitudes to 
be considered, the transverse ($A_{1/2}$)
and the longitudinal ($S_{1/2}$) amplitudes.
Those amplitudes are defined, 
at the resonance rest frame, as follows~\cite{Aznauryan12}:
\ba
A_{1/2} &= & 
\sqrt{\frac{2 \pi \alpha}{K}} \left< R, + \sfrac{1}{2} \right|
\varepsilon_+ \cdot J \left| N, - \sfrac{1}{2} \right>, 
\label{eqA12} \\
S_{1/2} &= & 
\sqrt{\frac{2 \pi \alpha}{K}} \left< R, + \sfrac{1}{2} \right|
\varepsilon_0 \cdot J \left| N, + \sfrac{1}{2} \right> \frac{|{\bf q}|}{Q},
\label{eqS12}
\ea
where $Q = \sqrt{Q^2}$ (assuming that $Q^2 > 0$),
as before $|{\bf q}|$ is the photon (and nucleon) momentum,
and $\varepsilon_\lambda$ ($\lambda=0,+$) 
is the photon polarization vector.
The momentum $|{\bf q}|$ is determined by
\ba
|{\bf q}| = \frac{\sqrt{Q_+^2 Q_-^2}}{2M_R},
\ea
where $Q_\pm^2 = (M_R \pm M)^2 + Q^2$.

\begin{figure}[t]
\vspace{.6cm}
\includegraphics[width=2.8in]{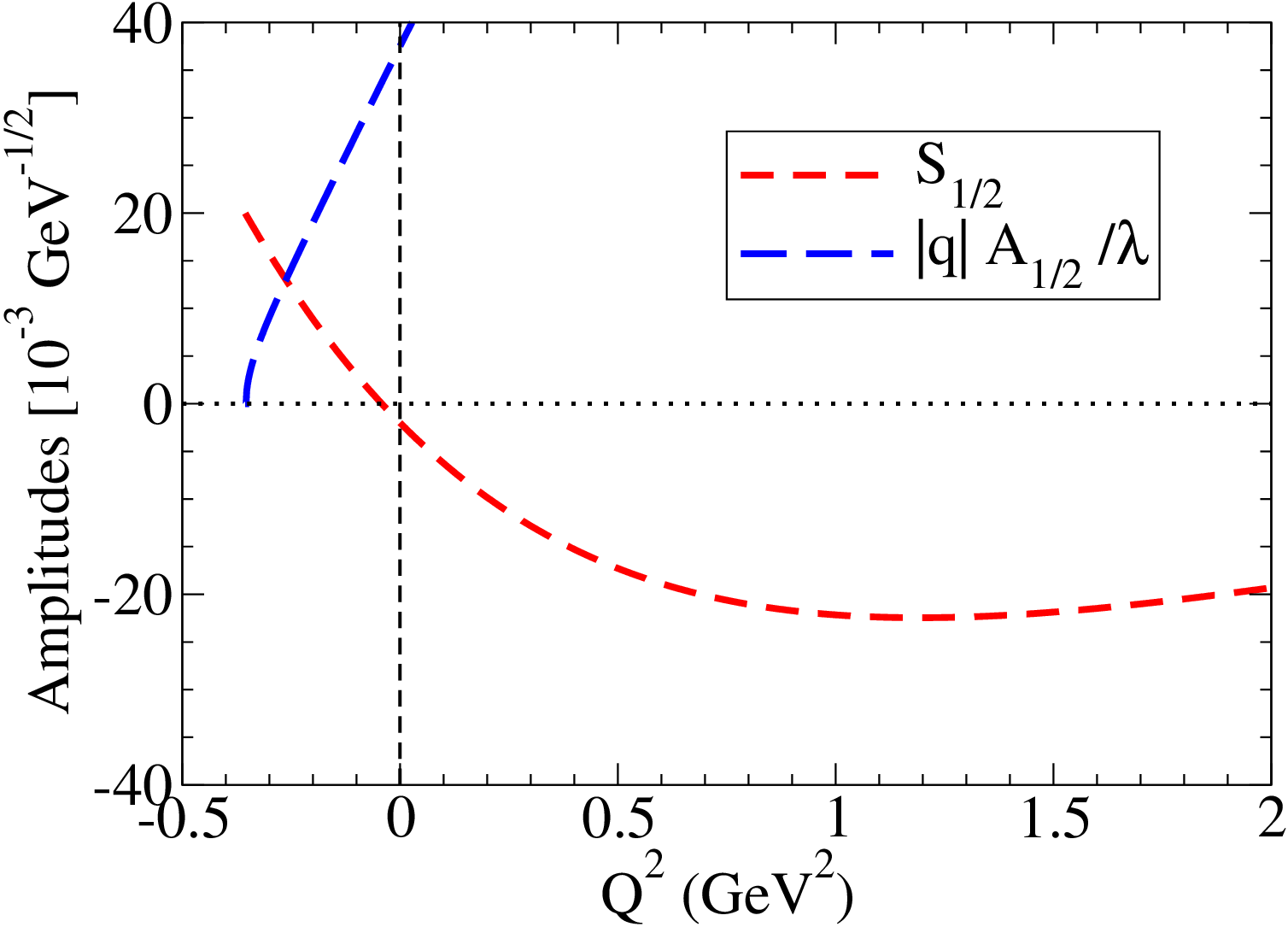}
\caption{\footnotesize
Comparison between the amplitude $S_{1/2}$ 
and $|{\bf q}| A_{1/2} / \lambda$, 
determined by the 
MAID2007 parametrization~\cite{Drechsel2007,Tiator16,MAID2009}.}
\label{figMAID}
\end{figure}

Based on the current (\ref{eqJmu}), we can write 
the amplitudes~\cite{Aznauryan12,N1535,S11scaling,Note}, as  
\ba
\hspace{-1cm}
A_{1/2} (Q^2)&=& 
2 b \tilde F_1 (Q^2), 
\label{eqA12new}
 \\
\hspace{-1cm}
S_{1/2} (Q^2)&=& 
- \sqrt{2} b (M_R -M) \frac{|{\bf q}|}{Q^2} \times \nonumber \\
& & 
 \left[\tilde F_1 (Q^2)  
- \frac{4 M_R^2 |{\bf q}|^2}{(M_R^2-M^2)Q_+^2} F_2(Q^2)
\right], 
\label{eqS12new}
\ea
where 
\mbox{
$b = e \sqrt{\frac{Q_+^2}{8 M (M_R^2-M^2)}}$},
as before, 
and $\tilde F_1$ is defined by Eq.~(\ref{eqF1til}).
The factor $e$ appears because 
the current $J^\mu$ is defined in units of the elementary electric charge.

In Eq.~(\ref{eqS12new}), we decompose 
the amplitude $S_{1/2}$ into a $\tilde F_1$ term 
and a term in $|{\bf q}|^2$, in order to facilitate 
the following discussion.

Based on Eqs.~(\ref{eqA12new})-(\ref{eqS12new}), 
we can conclude that if the term $|{\bf q}|^2 F_2$ can be dropped 
in comparison with $\tilde F_1$, 
we obtain immediately the Siegert's theorem condition, since 
\ba
A_{1/2} = 
2 b \tilde F_1 ,
\hspace{.7cm}
S_{1/2} = 
\sqrt{2} b\frac{|{\bf q}|}{M_R -M} \tilde F_1,
\label{eqSiegert2}
\ea
in the pseudo-threshold limit, $Q^2 \to - (M_R-M)^2$.

We look now for the results of the MAID2007 parametrization.
The results for the amplitude $S_{1/2}$ 
and $A_{1/2} |{\bf q}|/\lambda$  are presented in the 
Fig.~\ref{figMAID}. 
One can note in the figure, that  
$|{\bf q}| A_{1/2} \ne \lambda S_{1/2} $, 
since the functions differ at the pseudo-threshold, 
$Q^2 = Q_{PS}^2 \simeq - 0.36$ GeV$^2$,
when we start to draw the lines.

From  Eqs.~(\ref{eqA12new})-(\ref{eqS12new}), 
we can conjecture, that, the deviation from the 
Siegert's theorem condition (\ref{eqSiegert}) 
in the MAID2007 parametrization,
may be a consequence of the dependence on $|{\bf q}|$ of 
the function $F_2$, when $|{\bf q}| \to 0$.
Since we know from Eq.~(\ref{eqA12new}), 
that  $\tilde F_1 = {\cal O} (1)$
(because $\tilde F_1$ goes to a constant),
when $|{\bf q}| \to 0$, 
we may conjecture that $F_2 \propto 1/|{\bf q}|^3$,
in order to obtain 
$S_{1/2} = {\cal O} (1)$ in the MAID2007 parametrization.
In the conditions of the Siegert's theorem (\ref{eqSiegert}),
however, we expect $F_2 ={\cal O} (1/|{\bf q}|^{(2-n)} )$
with $n \ge 1$.

\subsection{Form factors}

We turn now for the analysis of the 
transition form factors.
The transition form factors $F_1$ and $F_2$ 
can be determined inverting  
Eqs.~(\ref{eqA12new})-(\ref{eqS12new}).
The results are
\ba
F_1 &=& \frac{1}{2b}\frac{ (M_R-M)^2 Q_+^2 }{ 4M_R^2 |{\bf q}|^2 } 
\left[ A_{1/2} - \lambda \frac{ S_{1/2} }{ |{\bf q}|} \right]  
\nonumber  \\
& &+  \frac{1}{2b} 
\left[ A_{1/2} - \lambda 
\frac{ S_{1/2} }{ |{\bf q}|} \right], 
\label{eqF1} \\
\eta F_2 & =& - \frac{1}{2b}
\frac{ (M_R-M)^2 Q_+^2 }{ 4M_R^2 |{\bf q}|^2 } 
\left[ A_{1/2} - \lambda \frac{ S_{1/2} }{ |{\bf q}|} \right]  \nonumber \\
& & + \frac{1}{2b}  \lambda \frac{ S_{1/2} }{ |{\bf q}|}.
\label{eqF2}
\ea
For the convenience of the discussion we multiply $F_2$ by $\eta$,
given by Eq.~(\ref{eqETA}).

From Eqs.~(\ref{eqF1})-(\ref{eqF2}), we can conclude, that, 
in the sum $\tilde F_1 =F_1 + \eta F_2 $,  all terms cancel, 
except for the term $A_{1/2}/(2b)$,
as expected from  Eq.~(\ref{eqA12new}).
From the equations, we can also conclude 
that if the factor ${\cal R}= A_{1/2} - \lambda S_{1/2} /|{\bf q}|$
does not vanish (${\cal R} \ne 0$), or 
it does not vanish fast enough with $|{\bf q}|$ 
when $|{\bf q}| \to 0$, then the form factors  $F_1$ 
and $\eta F_2$ diverge in the limit $|{\bf q}| \to 0$.

Considering the MAID2007 parametrization, where 
${\cal R}= {\cal O}(1/|{\bf q}|)$, since  $A_{1/2}, S_{1/2} = {\cal O} (1)$, 
we conclude that when $|{\bf q}| \to 0$,
$F_1, - \eta F_2 = {\cal O} (1/|{\bf q}|^3)$
(dominance of the term in $S_{1/2}$).
These results are consistent with the previous estimate of $F_2$
for the MAID2007 parametrization.
We checked numerically the divergence of 
the form factors $F_1$, $F_2$, in the 
MAID2007 parametrization.

If, however, the Siegert's theorem  (\ref{eqSiegert}) is valid, 
and ${\cal R}= {\cal O}( |{\bf q}|^n)$ with $n \ge 1$,
we conclude that
$F_1, - \eta F_2 = {\cal O} (1/|{\bf q}|^{(2-n)})$.
In the simplest case, when $n=1$, we 
obtain  $F_1, - \eta F_2 = {\cal O} (1/|{\bf q}|)$.
It is interesting to note, that, even in the 
conditions of the Siegert's theorem, the form 
factors $F_1, F_2$ may diverge in the pseudo-threshold limit.

We can show however, that, 
if we represent any of the functions $A_{1/2}$ and $S_{1/2}/|{\bf q}|$,
by a non-singular function $F$ of $Q^2$, we can write 
${\cal R}= {\cal O}( |{\bf q}|^n)$ with $n \ge 2$,
since in the expansion of a function $F(Q^2)$ 
in powers of $|{\bf q}|$, near $|{\bf q}|=0$, 
the first term vanishes. 
This result is the consequence of the relation 
$\frac{d F}{d |{\bf q}|} = \frac{4 M_R^2 |{\bf q}|}{M_R^2+ M^2 + Q^2} 
\frac{d F}{d Q^2}$, where $\frac{d F}{d |{\bf q}|}$ 
vanishes in the pseudo-threshold,
unless $\frac{d F}{d Q^2}$ diverges.
The implication of the previous result is that if 
${\cal R}= {\cal O}( |{\bf q}|^2)$, one obtains, 
according with the previous estimate, 
$F_1, -\eta F_2 = {\cal O}(1)$.
As consequence, 
both form factors $F_1$ and $F_2$, are 
finite at the pseudo-threshold. 
We present next a parametrization 
of the amplitudes $A_{1/2}$, $S_{1/2}$ 
consistent with the result  ${\cal R}= {\cal O}( |{\bf q}|^2)$.

\section{Modified MAID parametrization}

We consider now parametrizations of 
the  $\gamma^\ast N \to N(1535)$ helicity amplitudes,
that differs from the MAID2007 parametrization. 
Since the proposed parametrization is based 
in the form of the MAID2007 parametrization,
but is also compatible with the Siegert's theorem, 
we label it as MAID-SG parametrization (SG holds for Siegert).
In the MAID-SG parametrization one uses 
\ba
& &
\hspace{-.6cm}
A_{1/2} = a_0 \left( 1 + a_1 Q^2 \right) e^{ -a_4 Q^2},  
\label{eqA12SG} \\
& &
\hspace{-.6cm}
S_{1/2} = \frac{2 M_R |{\bf q}|}{Q_+^2}
s_0^\prime \left( 1 + s_1 Q^2 + s_2 Q^4\right) e^{ -s_4 Q^2} , 
\label{eqS12SG}
\ea
where the $a_0,a_1,a_4,s_1,s_2$ and $s_4$ 
are adjustable parameters 
and $s_0^\prime$ will be fixed by the Siegert's theorem 
condition~(\ref{eqSiegert}). 
Comparatively to the original 
MAID2007 parametrization~\cite{Drechsel2007,MAID2009}, 
we replaced $s_0 \to  (2 M_R |{\bf q}|) s_0^\prime /{Q_+^2}$
and add an extra term in $Q^4$ for $S_{1/2}$.
The extra term ($s_2 Q^4$) is important 
in order to obtain a parametrization 
based on small coefficients (between $10^{-3}$ and $10^3$),
in the spirit of the previous MAID parametrizations.
The factor $(2 M_R |{\bf q}|)/{Q_+^2}$ 
is included to give the correct behavior 
(proportional to $|{\bf q}|$) near $|{\bf q}|=0$, 
and preserve the high $Q^2$ 
behavior of the parametrization, since 
$2 M_R |{\bf q}|/{Q_+^2} = \sqrt{{Q_-^2}/{Q_+^2}} \to 1$, 
for very large $Q^2$.

\begin{table}[t]
\begin{center}
{\small
\begin{tabular}{l  c c c c}
\hline
\hline
 Amplitude $A_{1/2}$   
                  & $a_0$ & $a_1$ & $a_2$ & $a_4$  \\
\hline
MAID2007  & $66.40$ & $1.61$\sp & -- \sp & $0.70$ \\
MAID-SG    & $54.99$  & $2.09$\sp & -- \sp& $0.70$ \\
\hline
\hline
Amplitude $S_{1/2}$
      & $s_0$ & $s_1$ & $s_2$ & $s_4$ \\
\hline 
MAID2007  & $-2.00$ & $23.90$\sp & --\sp & $0.81$ \\
MAID-SG    & $-9.46 \,\frac{2M_R |{\bf q}|}{Q_+^2}$ & 
$11.57$\sp  &  $0.172$\sp & $0.93$ \\
\hline
\hline
\end{tabular}
}
\end{center}
\caption{Coefficients used in the calculation of the 
amplitudes $A_{1/2}$, $S_{1/2}$ 
based on the MAID2007 
and MAID-SG 
parametrizations.}
\label{tableS12}
\end{table}

Note that, using Eqs.~(\ref{eqA12SG}) and (\ref{eqS12SG}),
one has $A_{1/2} = {\cal O}(1)$ and $S_{1/2} = {\cal O}(|{\bf q}|)$,
when $|{\bf q}| \to 0$.
However, to ensure the 
Siegert's theorem, 
we still need to constrain the value of $s_0^\prime$
by Eq.~(\ref{eqSiegert}).
We fit all the coefficients to the MAID data~\cite{MAID2009}.
Since the MAID analysis gives negligible 
error bars for the amplitude $S_{1/2}$ when $Q^2 > 1.5$ GeV$^2$,
for the propose of the fit we use an error of 0.01$\times 10^{-3}$ GeV$^{-1/2}$.
The coefficients determined by the best fit are 
presented in Table \ref{tableS12}.

Although we could impose the Siegert's theorem 
refitting only the amplitude $S_{1/2}$, 
for a question of consistence one chose to 
fit both amplitudes simultaneously. 
The coefficients associated with the new fit 
based on Eqs.~(\ref{eqA12SG})-(\ref{eqS12SG})
are presented in Table \ref{tableS12},
in comparison with the MAID2007 parametrization,
which violates the Siegert's theorem. 
To facilitate the comparison with MAID2007, 
we replace  $s_0$ by $ (2 M_R |{\bf q}|) s_0^\prime/{Q_+^2} $.

\begin{figure}[t]
\vspace{.6cm}
\centerline{\mbox{
\includegraphics[width=2.8in]{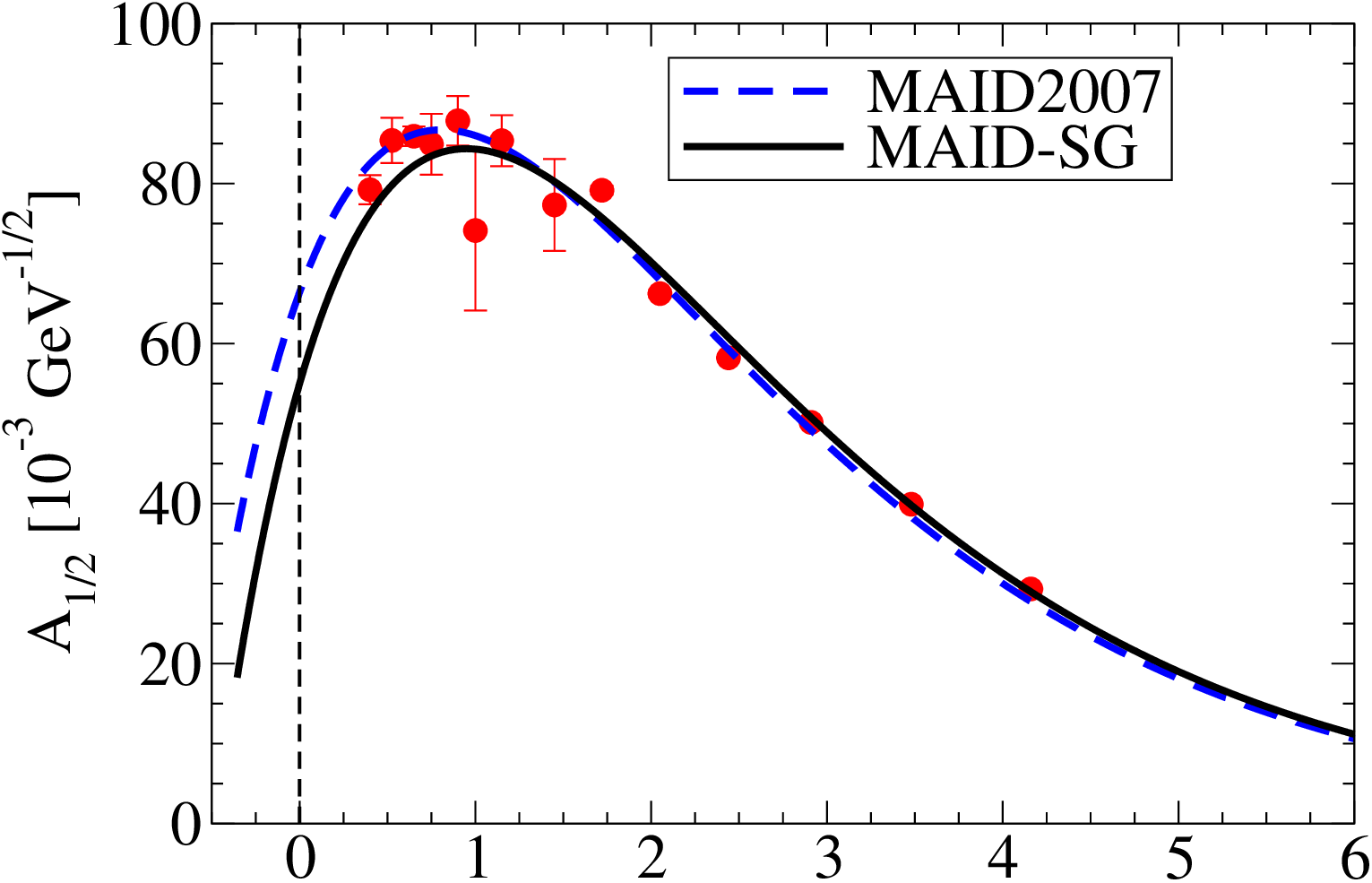}}}
\vspace{.8cm}
\centerline{\mbox{
\includegraphics[width=2.8in]{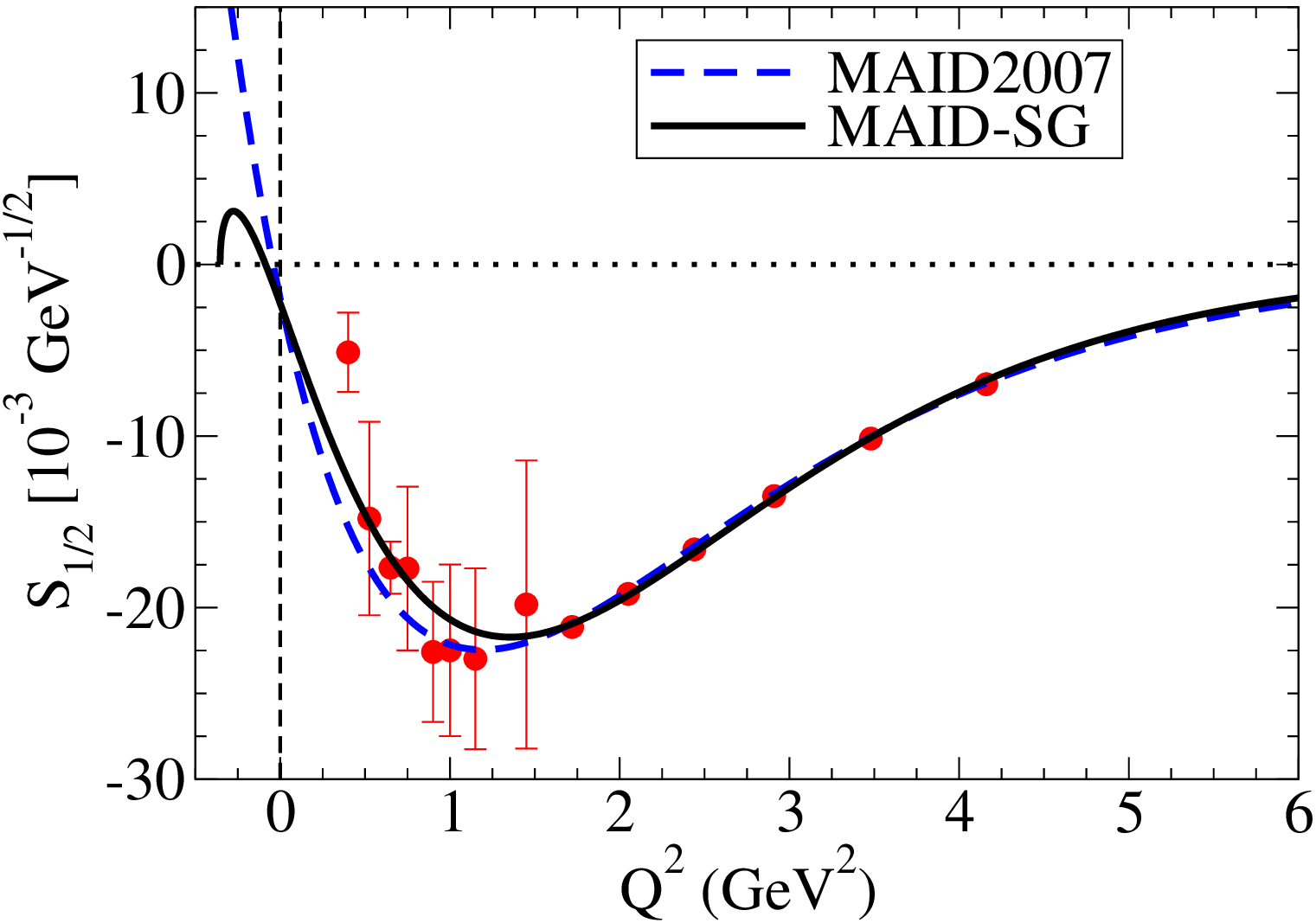}}}
\caption{\footnotesize
Amplitudes $A_{1/2}$ and $S_{1/2}$ determined by a fit 
consistent with the Siegert's theorem, compared 
with the MAID2007 parametrization. 
Data from the MAID analysis~\cite{MAID2009}.}
\label{figS12}
\end{figure}

\begin{figure}[t]
\vspace{.4cm}
\includegraphics[width=2.8in]{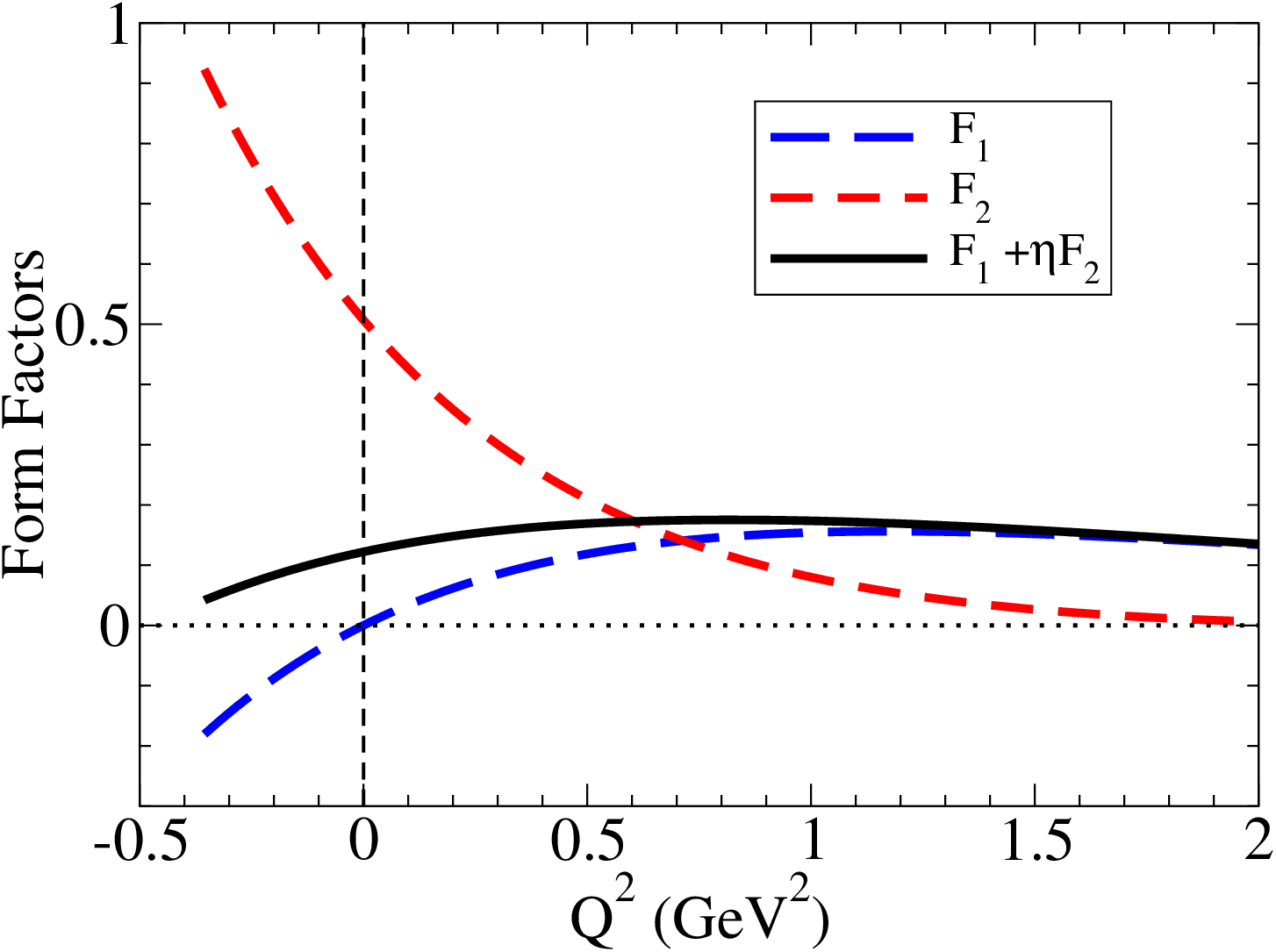}
\caption{\footnotesize
Form factors $F_1, F_2$ and $\tilde F_1$  determined by the 
MAID-SG parametrization.}
\label{figFF}
\end{figure}

The results for the amplitudes $A_{1/2}$ and 
$S_{1/2}$ in the MAID-SG parametrization are presented in Fig.~\ref{figS12}
(solid line),
and are compared with the result from MAID2007 (dashed line).
It is interesting to see that the   
two parametrizations are almost undistinguished  
for $Q^2 > 1.5$ GeV$^2$.
From the figure, we conclude, that, the constraints 
of the Siegert's theorem, can by included 
in the parametrization of the 
$\gamma^\ast N \to N(1535)$ helicity amplitudes, 
without a significant loss of accuracy.

The results for the amplitudes 
are consistent with the Siegert's theorem 
expressed in the form of Eq.~(\ref{eqSiegert2}), 
combined with $\tilde F_1 = {\cal O}(1)$.
Using the new parametrization 
for the amplitudes $A_{1/2}$ and $S_{1/2}$, 
it is possible now to look the form 
factors $F_1$ and $F_2$ 
based on Eqs.~(\ref{eqF1})-(\ref{eqF2}).
The results for the form factors are presented 
in the Fig.~\ref{figFF}.
In the figure, it is clear, 
that $F_1$ and $F_2$ are finite at the pseudo-threshold, 
as one expects from the dependence
${\cal R}= {\cal O}(|{\bf q}|^2)$, discussed previously.

We can calculate the explicit dependence of ${\cal R}$
near the  pseudo-threshold, 
using the functions $A,S$ defined by 
 $A \equiv A_{1/2}$  and 
$S_{1/2} \equiv  (2M_R |{\bf q}|) s_0^\prime/Q_+^2 S$.
One obtains then
\ba
{\cal R}= 
\frac{M_R}{M}
\left[
A' -  A \left( \frac{S'}{S} -\frac{1}{4M_R M}  \right)
\right] 
|{\bf q}|^2,
\label{eqR}
\ea
neglecting terms in ${\cal O}(|{\bf q}|^4)$.
In Eq.(\ref{eqR}), $A,S$ and  $A^\prime, S^\prime$ 
represent respectively the functions 
and the derivatives in the limit $Q^2=Q_{PS}^2$.

In Fig.~\ref{figFF}, one can also see, that the function 
$\tilde F_1$ is dominated by the form factors $F_1$,
for larger values of $Q^2$.
It is also possible to observe that 
the form factor $F_2$ has large values 
for $Q^2 < 0.5$ GeV$^2$, but decreases significantly 
for larger values of $Q^2$, and it is negligible 
for $Q^2 > 1.5$ GeV$^2$. 
A consequence of the result $F_2 \simeq 0$, 
is that the amplitudes $A_{1/2}$ and $S_{1/2}$ 
are correlated by the relation 
$S_{1/2} = - \frac{\sqrt{1 + \tau}}
{\sqrt{2}} \frac{M_R^2 - M^2}{2 M_R Q} A_{1/2}$,
where $\tau = \sfrac{Q^2}{(M_R + M)^2}$,
for $Q^2 > 1.5$ GeV$^2$~\cite{S11scaling}.
As discussed in Refs.~\cite{S11scaling,LambdaStar}
the result  $F_2 \simeq 0$,  suggests 
that there is a cancellation between
the valence quark contributions 
and the meson cloud contributions.

\section{Implication of the Siegert's theorem in other resonances}

\begin{figure}[t]
\vspace{.6cm}
\centerline{\mbox{
\includegraphics[width=2.6in]{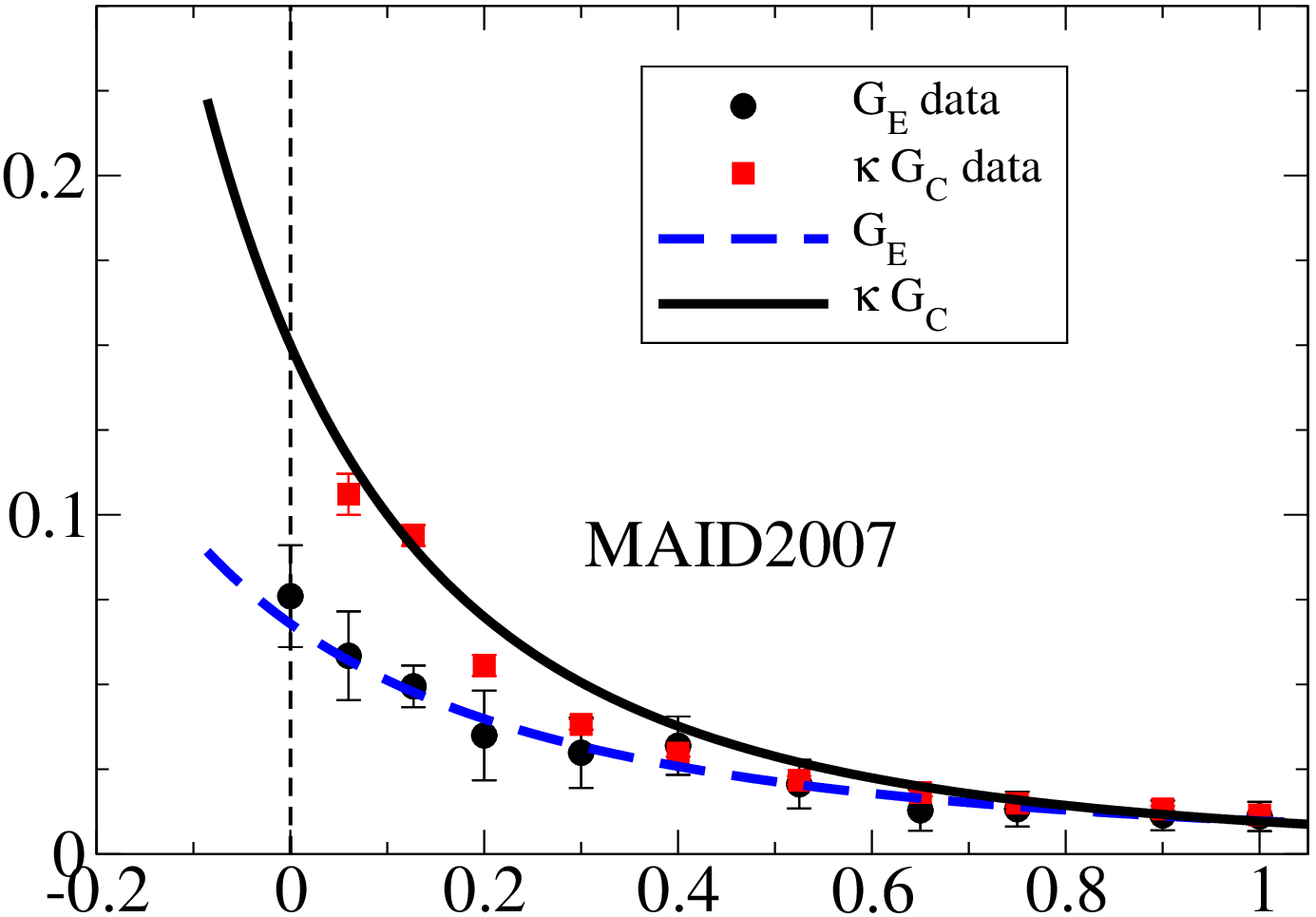}}}
\vspace{.8cm}
\centerline{\mbox{
\includegraphics[width=2.6in]{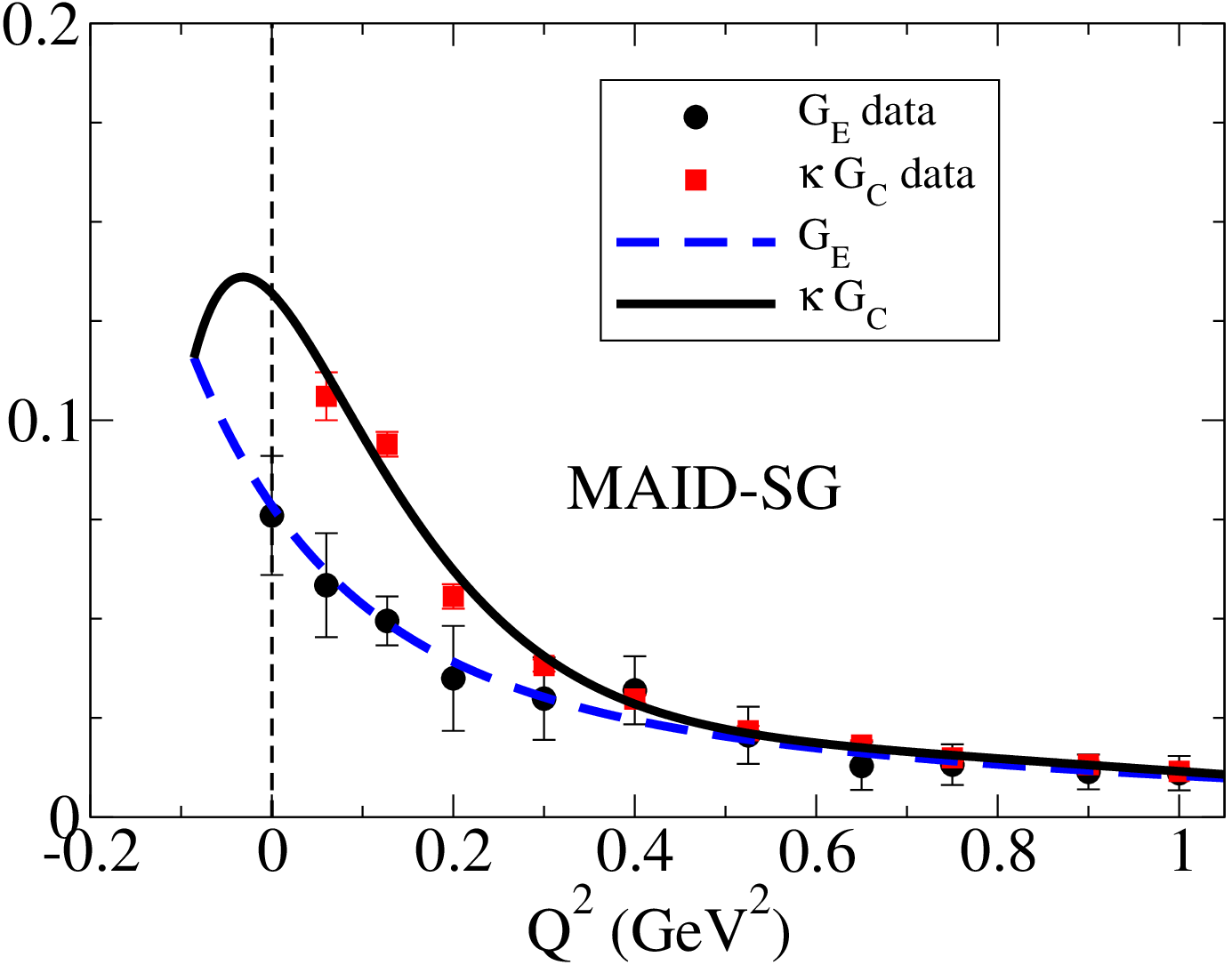}}}
\caption{\footnotesize
Electric and Coulomb quadrupole form factors 
for the $\gamma^\ast N \to \Delta(1232)$ transition.
At the top: MAID2007 parametrization~\cite{Drechsel2007}. 
At the bottom: improved parametrization 
consistent with the Siegert's theorem~\cite{newPaper}.
Data from Ref.~\cite{MokeevDatabase}. 
See details in Ref.~\cite{newPaper}.}
\label{figDelta}
\end{figure}

The constraints of the Siegert's theorem have implications
also in the helicity amplitudes associated with  
other $\gamma^\ast N \to N^\ast$ transitions.
In particular, the parametrization proposed here, 
can be used in the study 
of the $\gamma^\ast N \to N(1650)$ transition,
since it is also a $\frac{1}{2}^+ \to \frac{1}{2}^-$  transition.

In the case of the $\gamma^\ast N \to N(1520)$ transition 
the Siegert's theorem implies that 
$\frac{1}{2} E = \lambda_R S_{1/2}/|{\bf q}|$,
where $E \equiv -(A_{1/2} + \sqrt{3} A_{3/2} )$, 
is the electric amplitude in the transition, 
and  $\lambda_R = \sqrt{2}(M_R-M)$ 
($M_R$ is the resonance mass)~\cite{newPaper}.
One can see then, that
apart the  factor $1/2$ at the  l.h.s.,~and 
the replacement $A_{1/2} \to E$, 
the condition is the same as for the  
$\gamma^\ast N \to N(1535)$ transition.

Another interesting case is the $\gamma^\ast N \to \Delta(1232)$ transition. 
In this transition,
the electric ($G_E$) 
and the Coulomb ($G_M$) quadrupole form factors,
are related at  the pseudo-threshold limit, 
by the condition: $G_E = \kappa \, G_C$,
where $\kappa = \frac{M_\Delta -M}{2M_\Delta}$, and 
$M_\Delta$ is the $\Delta$ mass 
\cite{Devenish76,Jones73}.
When applied to the helicity amplitudes,   
one obtain the condition 
$E/|{\bf q}| = \lambda_\Delta S_{1/2}/|{\bf q}|^2$,
where $E \equiv A_{1/2} - A_{3/2}/\sqrt{3}$ 
is the electric amplitude and  $\lambda_\Delta = \sqrt{2}(M_\Delta -M)$.
One can show, that the previous condition 
for the amplitudes is violated by the MAID2007 parametrization.
Although the MAID2007 verify $E =  \lambda_\Delta S_{1/2}/|{\bf q}|$,
at pseudo-threshold 
(the r.h.s.~and the l.h.s.~vanish both),
this is not sufficient to ensure that 
$G_E = \kappa \, G_C$.  

In the Fig.~\ref{figDelta}, we compare at the top 
the form factors $G_E$ and $\kappa \, G_C$, given 
by the MAID2007 parametrization.
It is clear in the graph, that, 
the Siegert's theorem is violated.
At the bottom, we consider an improved parametrization 
where the Siegert's theorem is imposed 
and fitted to the  $G_E$ and $G_C$ data 
(defining a new MAID-SG parametrization). 
In this case, one can see the convergence of 
$G_E$ to $\kappa \, G_C$ at the pseudo-threshold.
The $\gamma^\ast N \to \Delta(1232)$ transition 
form factors and their relation with the 
Siegert's theorem are discussed in detail in Ref.~\cite{newPaper}.

\section{Summary and conclusions}

In the present article we discuss the 
implications of the constraints in the
$\gamma^\ast N \to N(1535)$ helicity amplitudes, 
when the nucleon and the 
resonance $N(1535)$ are both at rest (pseudo-threshold limit).
In this limit the 
transverse ($A_{1/2}$) and the longitudinal ($S_{1/2}$) 
amplitudes are related by the Siegert's theorem (\ref{eqSiegert}).
We concluded, that the Siegert's theorem is the consequence of the 
orthogonality between the nucleon and resonance states.

From the analysis of the structure of the current 
and the transition form factors, we conclude also, 
that, the amplitudes  $A_{1/2}$ and $S_{1/2}/|{\bf q}|$ 
are both finite and non-zero in the pseudo-threshold limit
[recall Eq.~(\ref{eqSiegert2}) with $\tilde F_1 = {\cal O}(1)$].
Based on this result, we explain why 
the MAID2007 parametrization for the amplitudes $A_{1/2}$ and $S_{1/2}$ 
violates the Siegert's theorem, 
and propose an alternative parametrization, 
consistent with both the  Siegert's theorem and the data.
The new parametrization is similar  
to the MAID2007 parametrization for 
both amplitudes when $Q^2 > 1.5$ GeV$^2$,
but deviates from MAID2007 for smaller values of $Q^2$.
In the new parametrization,
the amplitude $S_{1/2}$ differs more significantly 
from the MAID2007 parametrization for $Q^2 <0$,
and vanishes at the pseudo-threshold 
as expected ($S_{1/2} \propto |{\bf q}|$).

We concluded also, that, the Dirac and Pauli 
form factors are free of singularities 
at the pseudo-threshold as expected 
from the Siegert's theorem, expressed under the condition 
$A_{1/2} - \lambda S_{1/2}/|{\bf q}|= {\cal O} (|{\bf q}|^2)$,
near the pseudo-threshold.

The methods proposed in this article to study 
the structure of the helicity amplitudes and 
the structure of the transition form factors in the 
$\gamma^\ast N \to N(1535)$ transition, 
can be extended for the transitions $\gamma^\ast N \to \Delta(1232)$, 
$\gamma^\ast N \to N(1520)$ \cite{newPaper} and others.

\vspace{.2cm}
{\bf Acknowledgments:}
The author thanks Lothar Tiator for the 
useful discussions.
This work was supported by the Brazilian Ministry of Science,
Technology and Innovation (MCTI-Brazil).


\begin{thebibliography}{00}
\bibitem{Note} 
   Comparative to the Refs.~\cite{N1535,S11scaling} 
the form factors are modified by a sign.
\end{thebibliography}
\end{document}